\documentclass[english,aps,prb,twocolumn,showpacs,superscriptaddress]{revtex4-1}
\usepackage[T1]{fontenc}
\usepackage[latin9]{inputenc}
\usepackage{amsmath}
\usepackage{amssymb}
\usepackage{graphicx}
\usepackage{bm}
\usepackage{babel}
\usepackage{color}

\newcommand{\vk}{\mathbf{k}}
\newcommand{\vv}{\mathbf{v}}
\newcommand{\vl}{\mathbf{\Lambda}}
\newcommand{\vdl}{\delta\mathbf{\Lambda}}

\makeatletter

 \@ifundefined{textcolor}{}
 {
   \definecolor{BLACK}{gray}{0}
   \definecolor{WHITE}{gray}{1}
   \definecolor{RED}{rgb}{1,0,0}
   \definecolor{GREEN}{rgb}{0,1,0}
   \definecolor{DARKGREEN}{rgb}{0,0.5,0}
   \definecolor{BLUE}{rgb}{0,0,1}
   \definecolor{CYAN}{cmyk}{1,0,0,0}
   \definecolor{MAGENTA}{cmyk}{0,1,0,0}
   \definecolor{YELLOW}{cmyk}{0,0,1,0}
 }

\makeatother

\begin{document}

\title{Transport in multiband systems with hot spots on the Fermi
surface:\\ Forward-scattering corrections}

\author{Maxim Breitkreiz}
\email{maxim.breitkreiz@tu-dresden.de}
\affiliation{Institute of Theoretical Physics, Technische Universit\"at Dresden,
01062 Dresden, Germany}

\author{P. M. R. Brydon}
\affiliation{Condensed Matter Theory Center, Department of Physics,
University of Maryland, College Park, USA 20742}

\author{Carsten Timm}
\email{carsten.timm@tu-dresden.de}
\affiliation{Institute of Theoretical Physics, Technische Universit\"at Dresden,
01062 Dresden, Germany}

\date{June 6, 2014}

\begin{abstract}

Multiband models with hot spots are of current interest 
partly because of their relevance for the iron-based
  superconductors. In these materials,
the momentum-dependent scattering off spin fluctuations and the ellipticity of
the electron Fermi pockets are responsible for
anisotropy of the lifetimes of excitations 
around the Fermi surface. The deep minima of the lifetimes---the
so-called hot spots---have been assumed to contribute little to the
transport as is indeed predicted by a simple relaxation-time approach.
Calculating forward-scattering corrections to this approximation, we find that
the effective transport times are much more isotropic than the lifetimes and
that, therefore, the hot spots contribute to the transport even in
the case of strong spin-fluctuation scattering. We discuss this effect
on the basis of an analytical solution of the Boltzmann equation and calculate
numerically the temperature and doping dependence of the resistivity and
the Hall, Seebeck, and
Nernst coefficients.

\end{abstract}

\pacs{72.10.Di, 72.15.Lh, 74.70.Xa, 74.25.F-}

\maketitle

\section{Introduction}

Many materials of high current interest for condensed matter physics are metals
with strong spin fluctuations, for example doped cuprates and iron pnictides. In
both classes, spin fluctuations are thought to mediate the superconducting
pairing at relatively high temperatures.\cite{Johnston2010,Taillefer2010}
Spin fluctuations are also crucial in the normal state, where they
provide an important scattering mechanism and thus strongly affect transport.
The transport properties of the
pnictides are nevertheless quite distinct from the cuprates 
and show unusual temperature dependences.\cite{Zhu2008,Mun2009,Kondrat2009,%
Prelovsek2009,Matusiak2010,Eom2012,Ohgushi2012,Arsenievic2013,Blomberg2013} The
main ingredients needed for the description of transport in these systems have
been controversially
discussed.\cite{Fernandes2011,Kemper2011,Fanfarillo2012,Breitkreiz2013}

The scattering of electrons off spin fluctuations is governed by the spin
susceptibility. Close to an antiferromagnetic instability, the susceptibility
is strongly peaked in momentum space in the vicinity of the possible ordering
vectors $\mathbf{Q}$. Transport in such systems can thus often be understood
based on the concept of hot and cold regions of the Fermi
surfaces.\cite{Hlubina1995,Rosch1999} The hot regions are the parts of the Fermi
surfaces that are connected by the possible ordering vectors $\mathbf{Q}$. The
scattering is particularly strong in these regions. Conversely, in the cold
regions not connected by ordering vectors the scattering rate is lower.
If the difference in the scattering rate is large, i.e., close to the
instability, transport is thus dominated by the cold regions with high
conductivity, and the hot regions are then said to be
``short-circuited.''

The concept of hot and cold regions generally explains the experimental
observations for cuprates and was implicitly assumed to hold also for the
pnictides.\cite{Fernandes2011,Eom2012,Arsenievic2013,Blomberg2013} An analysis
of the lifetimes of excited electrons close to the Fermi surfaces seems to
support this picture,\cite{Kemper2011} with the imperfect nesting of
electron and hole Fermi pockets naturally leading to the appearance
of hot and cold regions with short and long lifetimes, respectively.

Within the \emph{relaxation-time approximation} (RTA), in which the
complex relaxation dynamics of each state is modeled by a simple exponential
decay, the transport relaxation time is approximated by the lifetime.
Since the conductivity is directly proportional to the
relaxation time, the states with short lifetimes then do not
contribute significantly to the transport.
In this paper we show that in multiband systems this effect can be compensated
if the forward-scattering corrections to the RTA are taken into account.

Forward-scattering corrections, which are equivalent to
vertex corrections in the Kubo formalism, have been studied extensively
for one-band models relevant for cuprates and
heavy-fermion systems.\cite{Kontani2008} The pnictides are, in contrast,
multiband systems with electron and hole Fermi
pockets. The study of two-band models with circular Fermi pockets has shown that
forward-scattering corrections to the RTA are huge
close to the antiferromagnetic instability and that they give rise to transport
anomalies such as a large enhancement of the Hall
coefficient\cite{Fanfarillo2012,Breitkreiz2013} and negative
magnetoresistance.\cite{Breitkreiz2013} The minority carriers, i.e., the
carriers on the smaller Fermi pocket, were found to exhibit negative transport
times, indicating a drift in the direction opposite of what one would expect
based on their charge. However, in the simplified models with circular
Fermi pockets all states on a
given Fermi pocket are equivalent because of rotational symmetry. They are thus
unable to address the concept of hot and cold regions, which 
only appear for noncircular Fermi pockets.

In this article we present a semiclassical Boltzmann theory of transport for a
two-band model with elliptical electron pockets relevant for the iron
pnictides. We show that due to the forward scattering, the hot-spot picture fails
for the pnictides even for very strong spin fluctuations and highly elliptical
electron pockets. In contrast to the lifetimes, which are
highly anisotropic around the Fermi pockets with deep minima at
the hot spots, the effective transport relaxation times are
found to be much more isotropic and to show no special features at the
hot spots. Our approximate analytical solution of the Boltzmann equation
provides insight into the mechanism behind this effect: The anisotropy of
the spin-fluctuation scattering extends the effective relaxation
time. At the hot spots, the reduction of the relaxation
time due to the stronger scattering is thus compensated by the extension
due to the higher anisotropy.

To elucidate the consequences of this mechanism, we calculate numerically the
temperature-dependent transport coefficients from the full Boltzmann
equation and compare them to the analytical solution and the
RTA, finding that the RTA makes qualitatively incorrect predictions.
For strongly momentum-dependent scattering, we find large transport anomalies
as well as a strong doping dependence.

The remainder of this paper is organized as follows: In Secs.\ \ref{model} and
\ref{BF}, we present the two-band model, give expressions for the scattering
rates, and set up the Boltzmann equation for our model. To gain insight into
the physics, we present in Sec.\ \ref{AA} an analytical solution to
leading order in the ellipticities of the electron pockets. Higher-order 
corrections are discussed in the appendix. In Sec.\ \ref{NR}, we present
full numerical solutions of the Boltzmann equations. We also calculate the
temperature dependence of the resistivity and the Hall, Seebeck, and Nernst
coefficients. Finally, we draw some conclusions in Sec.\ \ref{Con}.

\section{Model}
\label{model}

\begin{figure}[t]
\includegraphics[width=\columnwidth]{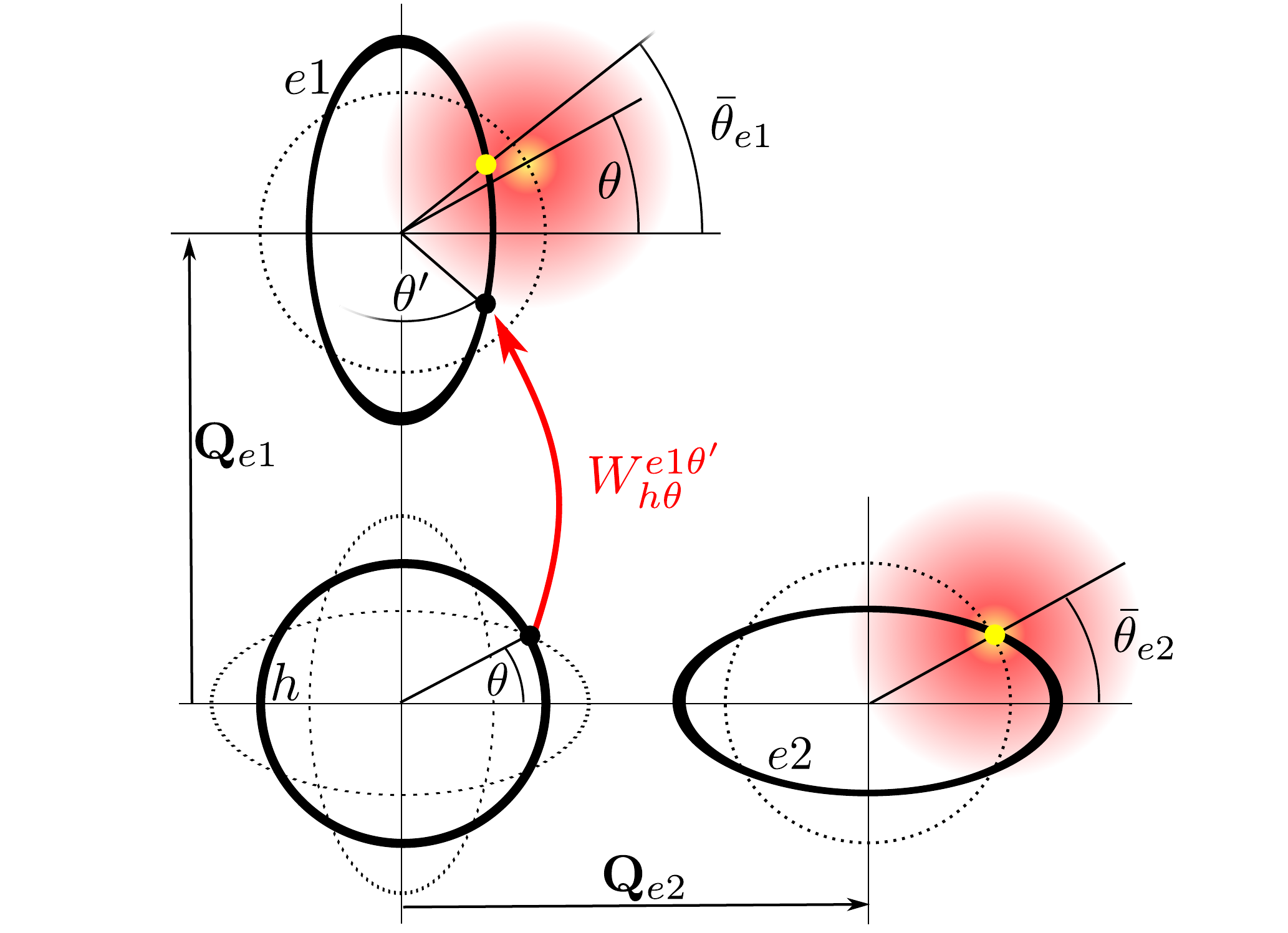}
\caption{(Color online) Illustration of the Fermi pockets and the scattering
rates. An electron in state $|h,\theta\rangle$ is scattered to
$|e1,\theta'\rangle$. The yellow (light gray) dots indicate the maxima of the
scattering rates $W_{h\,\theta}^{e1\,\theta'}$ and $W_{h\,\theta}^{e2\,\theta'}$
as functions of the polar angle $\theta'$ on the target Fermi pocket. The maxima
stem from the enhanced spin susceptibility (color gradient) for the scattering
wave vectors $\mathbf{Q}_{e1}$ and $\mathbf{Q}_{e2}$. The thin
dotted lines show the Fermi surfaces displaced by the nesting vectors.
The hot spots are located at the resulting crossing points.}
\label{fig:figure1}
\end{figure}

We model the FeAs layers of the iron pnictides by an effective two-dimensional
two-band model with the dispersions in the single-iron unit
cell~\cite{Johnston2010} given by
\cite{Brydon2011}
\begin{eqnarray}
\varepsilon_{h\vk}&=&\varepsilon_{h}-\mu+2t_{h}\,(\cos k_{x}a+\cos k_{y}a),  \\
\varepsilon_{e\vk}&=&\varepsilon_{e}-\mu+t_{e,1}\cos k_{x}a\,\cos k_{y}a
\nonumber \\
&& {}- t_{e,2}\,\xi\,(\cos k_{x}a+\cos k_{y}a),
\label{eek}
\end{eqnarray}
where $a$ is the iron-iron separation. As illustrated in
Fig.\ \ref{fig:figure1}, 
the band $h$ gives rise to a nearly circular hole Fermi pocket at the center
of the Brillouin zone, while the band $e$ forms two electron
pockets $e1$ 
and $e2$, displaced by $\mathbf{Q}_{e1}=(0,\pi/a)$ and 
  $\mathbf{Q}_{e2}=(\pi/a,0)$,
respectively. The parameter $\xi$ controls the ellipticity of the
electron pockets. The chemical potential $\mu$ is
determined by the filling $n$, i.e., the number of electrons per unit
cell, which can be tuned by doping in the pnictides. The
  filling $n$ determines the sizes of the Fermi pockets. For 
$n\approx2.08$ the areas of the three pockets are nearly equal,
while for smaller (larger) $n$ the hole pocket (electron pockets) become larger.
Following Ref.\ \onlinecite{Brydon2011}, we take
$\varepsilon_h=-3.5\,t_h$, $\varepsilon_e=3\,t_h$, $t_{e,1}=4\,t_h$ and
$t_{e,2}=t_h$.

It is widely accepted that repulsive interactions between the
nested electron and hole pockets drive a magnetic
instability towards a stripe spin-density wave in the pnictide parent
compounds with magnetic ordering vector ${\bf Q}_{e1}$ or
${\bf Q}_{e2}$.~\cite{EreChu2010} Above the magnetic
transition temperature, we 
therefore expect that the spin susceptibility will display
pronounced peaks at these vectors. Because of the ellipticity of the
electron pockets, however, the nesting is imperfect and distinct hot
spots develop at the points on the electron and hole Fermi pockets
separated by ${\bf Q}_{e1}$ or ${\bf Q}_{e2}$, see
Fig.~\ref{fig:figure1}. The positions of the hot spots change with
the doping:~\cite{Brydon2011,Fernandes2011} for underdoping
($n<2.08$) the hot spots 
are located near the major axis of the electron pockets, while at
overdoping ($n>2.08$) the hots spots shift to the minor
axis. On the hole pocket, the hot spots shift from the axes to the
diagonal and back again as one dopes across the
antiferromagnetic dome.

We assume that the transport behavior is dominated by the scattering off spin
fluctuations, which we model by the phenomenological susceptibility
proposed by Millis, Monien, and Pines,\cite{Millis1990} with
temperature-dependent parameters based on neutron-scattering
experiments.\cite{Inosov2010} Although this ignores the
anisotropy of the magnetic excitations in the pnictides caused by the
ellipticity of the electron Fermi pockets,\cite{Diallo2010} we
shall see that the precise form of the susceptibility is less
important for the transport than the anisotropy of the scattering rate.
Together with momentum-independent impurity scattering, the scattering rate
from a single-electron state $|b, \vk\rangle$ to a state $|b', \vk'\rangle$,
where $b=e$, $h$ denotes the band, can be written as\cite{Rice1967}
\begin{eqnarray}
W_{b\vk}^{b'\vk'} &=& (1-\delta_{bb'})\,
  W_{\text{sf}}\, \frac{p_T(\varepsilon_{b\vk}-\varepsilon_{b'\vk'})}
    {\left(\varepsilon_{b\vk}-\varepsilon_{b'\vk'}\right)^{2}
     + \omega_{\vk,\vk'}^{2}} \nonumber \\
&& {}+ \delta(\varepsilon_{b\vk}-\varepsilon_{b'\vk'})\, W_{\text{imp}},
\label{eq:01}
 \end{eqnarray}
where $W_{\text{sf}}$ and $W_{\text{imp}}$ represent the overall
strength of the scattering off spin fluctuations and impurities, respectively,
$p_T(x)\equiv x\left(\coth x/2k_BT-\tanh x/2k_BT\right)$, and
\begin{equation}
\omega_{\vk,\vk'} \equiv \Gamma_{T}\, \Big( 1
  +\xi_{T}^{2} \min_{\mathbf{Q}}[(\vk-\vk'+\mathbf{Q})^{2}]\Big),
\end{equation}
where the four possible values for $\mathbf{Q}$ are $\pm\mathbf{Q}_{e1}$ and
$\pm\mathbf{Q}_{e2}$. With the Curie-Wei\ss{} temperature
$-\theta_\mathrm{CW}<0$, the frequency scale and the correlation
length are given by \cite{Fanfarillo2012,Inosov2010}
$\Gamma_T=\Gamma_0\,(T+\theta_\mathrm{CW})/\theta_\mathrm{CW}$ and
$\xi_T=\xi_0\,\sqrt{\theta_\mathrm{CW}/(T+\theta_\mathrm{CW})}\:\exp(-T/T_0)$,
respectively. Following Ref.\ \onlinecite{Fanfarillo2012}, we here introduce an
additional exponential decay of $\xi_T$ to account for the high-temperature
behavior and choose $T_0=200\,\mathrm{K}$. Following Ref.\
\onlinecite{Inosov2010}, we take $\xi_0=10\,a$,
$\theta_\mathrm{CW}=30\,\mathrm{K}$ and
$\Gamma_0=4.2\,\mathrm{meV}$.
The resulting form of $\omega_{\vk,\vk'}$ and thus $W_{b\vk}^{b'\vk'}$ is only
valid as long as the system does not order antiferromagnetically or becomes
superconducting.

The transport is governed by states on the Fermi pockets, denoted by
$|s,\theta\rangle$, where $s=h$, $e1$, $e2$ is the pocket index and $\theta$
is the polar angle along the pocket, see
Fig.\ \ref{fig:figure1}. From Eq.\ (\ref{eq:01}) we see that in the
low-temperature regime, $k_BT\ll\varepsilon_F$, the scattering rate is sharply
peaked at $\varepsilon_{b\vk}=\varepsilon_{b'\vk'}$ so that scattering is
nearly elastic. We exploit this fact by writing
\begin{equation}
W_{b\vk_{F}}^{b'\vk'}
  \approx \delta(\varepsilon_{b'\vk'}-\varepsilon_{F})\,
  W_{s\theta}^{s'\theta'},
\label{eq:elastic}
\end{equation}
where
\begin{equation}
W_{s\theta}^{s'\theta'} \equiv (1-\delta_{bb'})\, W_{\text{sf}}\int
  d\varepsilon\,
  \frac{p_T(\varepsilon)}{\varepsilon^{2}+\omega_{\vk,\vk'}^{2}}
  + W_{\text{imp}}
\label{eq:02}
\end{equation}
is the effective elastic scattering rate between states on the Fermi pockets
$s$, $s'$ belonging to the bands $b$, $b'$.
Since the spin susceptibility and thus $W_{b\vk}^{b'\vk'}$ is strongly momentum
dependent, the elastic scattering rate $W_{s\theta}^{s'\theta'}$ strongly
depends on the angles $\theta$ and $\theta'$, in particular on the change in
angle, $\theta'-\theta$. This is what we call \emph{anisotropic} scattering in
the following.

More specifically, the scattering anisotropy stems from the
peaks in the spin susceptibility at 
the wave vectors $\pm\mathbf{Q}_{e1}$ and $\pm\mathbf{Q}_{e2}$. For an initial
state $|h,\theta\rangle$ with wave vector $\vk$, the scattering rate has maxima
for the final states $|e1,\bar{\theta}_{e1}\rangle$ and
$|e2,\bar{\theta}_{e2}\rangle$, defined as the states on the Fermi pockets
$e1$, $e2$ with wave vectors closest to $\vk+\mathbf{Q}_{e1}$ and
$\vk+\mathbf{Q}_{e2}$, respectively, see Fig.\ \ref{fig:figure1}.
Similarly, for an initial state $|e1, \theta\rangle$ ($|e2, \theta\rangle$) with
wave vector $\vk$, the scattering rate has a maximum for the final state
$|h,\bar{\theta}_{h}\rangle$ with wave vector closest to $\vk-\mathbf{Q}_{e1}$
($\vk-\mathbf{Q}_{e2}$), where $\bar{\theta}_{h}\approx\theta$ since the hole
pocket is nearly circular.

The scattering rate summed over all final states determines the characteristic
lifetime of the state $|s,\theta\rangle$,
\begin{equation}
\tau_{s\theta}=\bigg(\frac{1}{2\pi}\sum_{s'}\int
  d\theta'\, N_{s'\theta'}\, W_{s\theta}^{s'\theta'}\bigg)^{\!-1},
\label{eq:lt}
\end{equation}
where $N_{s\theta}=|d\vk_{F,s\theta}/d\theta|/\pi \hbar |\vv_{F,s\theta}|$  is
the density of states, with the spin degeneracy included, of pocket $s$ at the
polar angle $\theta$ and $\vk_{F,s\theta}$ and $\vv_{F,s\theta}$ are the Fermi
momentum and the Fermi velocity, respectively. In contrast to the
transport relaxation time, which
will be discussed below, the lifetime only depends on the integrated
scattering strength and is independent of the precise shape of
$W_{s\theta}^{s'\theta'}$ as a function of $\theta'$.

\section{Boltzmann formalism}
\label{BF}

Our starting point is the semiclassical Boltzmann transport equation for a
multiband system,
\begin{eqnarray}
\lefteqn{-f'_0(\varepsilon_{b\vk})\, \mathbf{E}\cdot\vv_{b\vk}
  -\frac{e}{\hbar}\, \mathbf{B}
  \cdot(\vv_{b\vk}\times\nabla_{\vk})\, g_{b\vk} } \nonumber \\
&& \qquad = \sum_{b'\vk'} W_{b\vk}^{b'\vk'}\,
  (g_{b\vk}-g_{b'\vk'}), \hspace{4em}
\label{Boltzmanneq}
\end{eqnarray}
where $\mathbf{E}=(E_x,E_y,0)$ and $\mathbf{B}=(0,0,B)$ are weak uniform
electric and magnetic fields, respectively, $\mathbf{v}_{b\vk} \equiv
\hbar^{-1}\,\nabla_{\vk}\varepsilon_{b\vk}$ is the velocity and
$g_{b\vk}\equiv f_{b\vk}-f_0(\varepsilon_{b\vk})$ is the difference between the
non-equilibrium distribution function $f_{b\vk}$ and the Fermi-Dirac
distribution $f_0(\varepsilon_{b\vk})$. This difference is of the general
form\cite{Sondheimer1962,Taylor1963,Beenakker2011}
\begin{equation}
g_{b\vk}=-f'_0(\varepsilon_{b\vk})\,
  \mathbf{E}\cdot(\vl_{b\vk}+\vdl_{b\vk}),
\label{eq:ans}
\end{equation}
with the as yet unknown vector mean free path $\vl_{b\vk}+\vdl_{b\vk}$. Here,
$\vl_{b\vk}$ ($\vdl_{b\vk}$) is of zero (first) order in the magnetic field
$\mathbf{B}$. For states on the Fermi pockets we write $\vl_{s\theta}$,
$\vdl_{s\theta}$ with obvious definitions.

Inserting Eqs.\ (\ref{eq:elastic}), (\ref{eq:02}), and (\ref{eq:ans}) into the
Boltzmann equation (\ref{Boltzmanneq}) and using $\sum_{b'\vk'}
=\sum_{s'}\int \frac{d\theta'}{2\pi}N_{s'\theta'}\int d\varepsilon_{b'\vk'}$,
one finds for states at the Fermi
energy\cite{Beenakker2011}
\begin{eqnarray}
\vl_{s\theta} &=& \tau_{s\theta}\, \mathbf{v}_{s\theta}
  + \tau_{s\theta}\sum_{s'}\int \frac{d\theta'}{2\pi}\,
  N_{s'\theta'}\,W_{s\theta}^{s'\theta'}\,\vl_{s'\theta'},
\label{eq:03a} \\
\vdl_{s\theta} &=& \tau_{s\theta}\, \eta_s\,
  \frac{eB}{\pi\hbar^2}\, \frac{1}{N_{s\theta}}\,
  \frac{\partial\vl_{s\theta}}{\partial\theta} \nonumber \\
&& {} + \tau_{s\theta}\sum_{s'}\int \frac{d\theta'}{2\pi}\,
  N_{s'\theta'}\,W_{s\theta}^{s'\theta'}\,\vdl_{s'\theta'} ,\hspace{4em}
\label{eq:03b}
\end{eqnarray}
where $\eta_h=1$ and $\eta_{e1}=\eta_{e2}=-1$. The RTA consists of
neglecting the forward-scattering corrections in Eqs.\
(\ref{eq:03a}) and (\ref{eq:03b}), i.e., the second terms on the right-hand sides.
Thus in the RTA one obtains
\begin{eqnarray}
\vl_{s\theta} &=& \vl_{s\theta}^{(0)} \;\equiv\;
  \tau_{s\theta}\, \mathbf{v}_{s\theta},
\label{eq:04a} \\
\vdl_{s\theta} &=& \vdl_{s\theta}^{(0)} \;\equiv\;
  \tau_{s\theta}\, \eta_s\,
  \frac{eB}{\pi\hbar^2}\, \frac{1}{N_{s\theta}}\,
  \frac{\partial\vl_{s\theta}^{(0)}}{\partial\theta}.
\label{eq:04b}
\end{eqnarray}
Evidently, within the RTA
the solution is determined by the bare lifetimes $\tau_{s\theta}$ given in Eq.\
(\ref{eq:lt}). The RTA becomes exact if the scattering rate is isotropic around
the Fermi pockets so that the forward-scattering corrections average out. For a
nonzero scattering anisotropy, however, the result may differ significantly from the
RTA.\cite{Breitkreiz2013}

The charge current $\mathbf{J}=\sigma\mathbf{E}$
is controlled by the conductivity tensor $\sigma$,
which is in turn determined by the vector mean free
path,\cite{Beenakker2011}
\begin{equation}
\sigma^{ij} = e^2 \sum_{s} \int\frac{d\theta}{2\pi}\,
  N_{s\theta}\, v_{s\theta}^{i}\,
  \big(\Lambda_{s\theta}^{j}+\delta\Lambda_{s\theta}^{j}\big)
  \equiv \sum_{s} \int\frac{d\theta}{2\pi}\,\sigma^{ij}_{s\theta}.
\label{eq:sigma}
\end{equation}
Writing $\mathbf{E}=E\,(\cos\phi,\sin\phi,0)$, we find the current parallel to
the electric field as
\begin{eqnarray}
\frac{\mathbf{J}\cdot\mathbf{E}}{E} &=& \sum_{s} \int\frac{d\theta}{2\pi}\, \big(
  \sigma_{s\theta}^{xx}\cos^2\phi
  + \sigma_{s\theta}^{yy}\sin^2\phi \nonumber \\
&& {} + \sigma_{s\theta}^{xy}\cos\phi\sin\phi
  + \sigma_{s\theta}^{yx}\cos\phi\sin\phi \big)
  \nonumber \\
&\equiv& \sum_{s} \int\frac{d\theta}{2\pi}\,J_{s\theta},
\label{eq:current12}
\end{eqnarray}
where $J_{s\theta}$ is the contribution of the state $|s,\theta\rangle$
to the current.

\section{Analytical results}
\label{AA}

To gain insight into transport beyond the RTA, we now construct an approximate
analytical solution of Eqs.\ (\ref{eq:03a}) and (\ref{eq:03b}) that fully
accounts for the anisotropic scattering. We will first discuss a few reasonable
assumptions that make an analytical solution feasible. The full numerical
solution is discussed in Sec.~\ref{NR}.

As illustrated in Fig.\ \ref{fig:figure1}, the scattering rate
$W_{s\theta}^{s'\theta'}$ understood as a function of $\theta'$ has a maximum at
$\theta'=\bar{\theta}_{s'}$, which of course depends on $\theta$. The small
difference between $\theta$ and $\bar{\theta}_{s'}$ stems from the ellipticity
of the electron pockets. We now make two simplifying assumptions:
(\textit{i}) The peak of the scattering rate $W_{s\theta}^{s'\theta'}$ as a
function of $\theta'$ is assumed to be symmetric around
$\theta'=\bar{\theta}_{s'}$, and (\textit{ii}) the peak width is small on the
scale on which the Fermi velocity $|\vv_{s\theta}|$ and the density of states
$N_{s\theta}$ vary. Both assumptions become exact in the limit of very strongly
peaked spin susceptibility, i.e., as the magnetic instability is approached. In
the opposite limit of isotropic scattering, the forward-scattering corrections
cancel out so that we also obtain the exact results.

On the right-hand side of Eq.\ (\ref{eq:03a}), we split
$\vl_{s'\theta'}$ into contributions parallel and perpendicular to
$\vl_{s'\bar\theta_{s'}}$,
\begin{equation}
\vl_{s'\theta'} = \frac{|\vl_{s'\theta'}|}{|\vl_{s'\bar{\theta}_{s'}}|}
  \big[\vl_{s'\bar{\theta}_{s'}}
  \cos(\theta'-\bar{\theta}_{s'})
  + \hat{\mathbf{z}}\times\vl_{s'\bar{\theta}_{s'}}
  \sin(\theta'-\bar{\theta}_{s'})\big] .
\end{equation}
By virtue of the assumptions (\textit{i}) and (\textit{ii}), the sine term
drops out and we obtain
\begin{equation}
\vl_{s\theta} = \vl_{s\theta}^{(0)}
  + \left(1-\frac{1}{2}\, \delta_{s,h}\right)
  \sum_{s'} a_{s\theta}^{s'}\, \vl_{s'\bar{\theta}_{s'}},
\label{eq:05}
\end{equation}
where 
\begin{equation}
a_{s\theta}^{s'} \equiv (1+\delta_{s,h})\,
  \tau_{s\theta} \int \frac{d\theta'}{2\pi}\, N_{s'\theta'}\,
  W_{s\theta}^{s'\theta'} \cos(\theta'-\bar{\theta}_{s'})
\label{eq:06}
\end{equation}
parametrizes the scattering anisotropy and in the following will be
referred to as the \emph{anisotropy parameter}.
The Kronecker symbols $\delta_{s,h}$ appearing in Eqs.\
(\ref{eq:05}) and (\ref{eq:06}) ensure that $a_{s\theta}^{s'}\in[0,1]$ and that
$a_{s\theta}^{s'}\rightarrow1$ corresponds to the limit of strong scattering anisotropy,
$W_{s\theta}^{s'\theta'}\propto\delta(\theta'-\bar{\theta}_{s'})$, while
$a_{s\theta}^{s'}\rightarrow0$ gives the case of isotropic scattering, where the
RTA result is recovered.

\begin{figure}[t]
\includegraphics[width=\columnwidth]{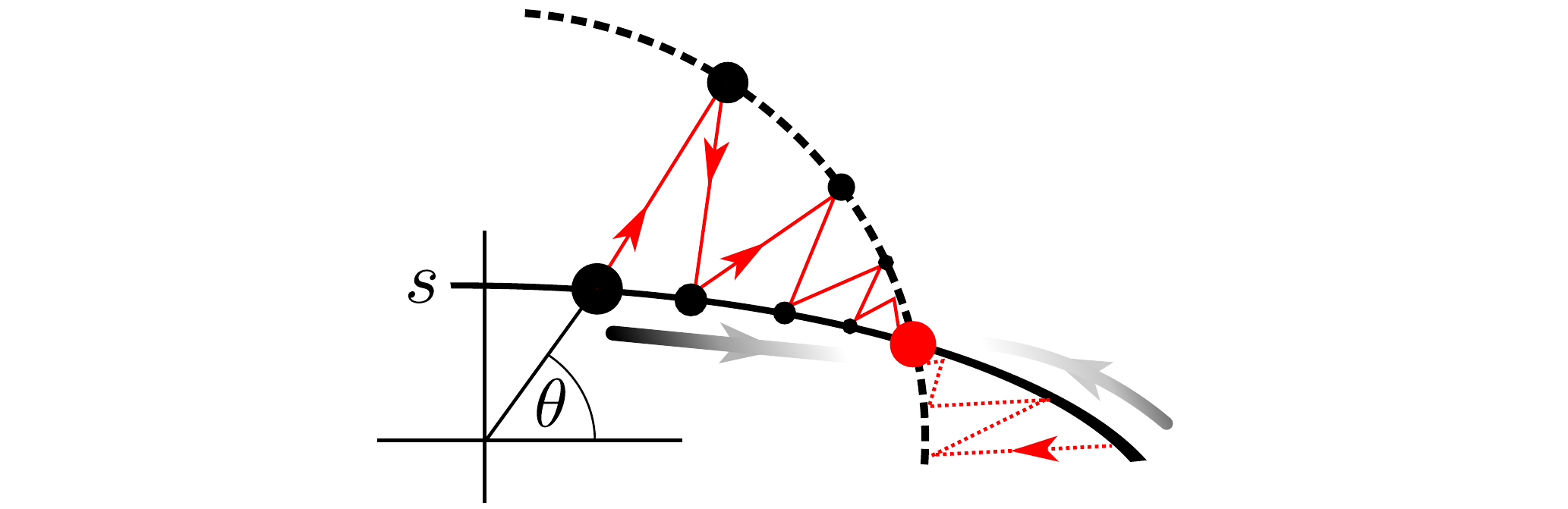}
\caption{(Color online) Sketch of multiple scattering. During the process, an
electron initially in state $|s,\theta\rangle$ effectively scatters between
Fermi pockets towards the closest hot spot (red/gray dot). The sequence
of states (black dots) is given by the maximum of the scattering rate. Their
decreasing contribution to the vector mean free path $\vl$ of the
original state $|s,\theta\rangle$ is indicated by the decreasing size of the
dots.}
\label{fig:figure2}
\end{figure}

Iterating Eq.\ (\ref{eq:05}), we obtain $\vl$ in terms of
$\vl^{(0)}$ as a power series in the anisotropy parameter.
We now discuss the states appearing in this series.
The zero-order contribution to $\vl_{s\theta}$ is of course
$\vl_{s\theta}^{(0)}$, the RTA result for the same state $|s,\theta\rangle$.
The first-order term involves $\vl_{s'\bar\theta_{s'}}^{(0)}$ for the
state $|s',\bar\theta_{s'}\rangle$. This is the \emph{final} state on the Fermi
pocket $s'\neq s$ to which the \emph{initial} state $|s,\theta\rangle$ has the
largest scattering rate. Due to the ellipticity of the electron
pockets, the shift of the angle, $\bar{\theta}_{s'}-\theta$, is always
directed towards the closest hot spot, i.e., the intersection of the Fermi
pocket $s$ with pocket $s'$ shifted by the appropriate vector $\mathbf{Q}$.
The state appearing in the second-order term is the one reached from
$|s',\bar\theta_{s'}\rangle$ with the largest scattering rate, again shifted
towards the closest hot spot. The states appearing in all higher-order terms
are obtained in the same way. The whole process can be interpreted as an
effective hopping of the electron along a sequence of states, as illustrated by
Fig.\ \ref{fig:figure2}.

The contribution to $\vl_{s\theta}$ from
$\vl_{s_\nu\theta_\nu}^{(0)}$ of the
state $|s_\nu,\theta_\nu\rangle$ reached after $\nu$ hopping
events involves the product of $\nu$ anisotropy parameters at
$\theta$, $\theta_1$, $\ldots$, $\theta_{\nu-1}$. Since the angular shift
between successive hopping events is due to the ellipticity of the
electron pockets, and vanishes for a purely circular pocket, it is
small for small ellipticities. Indeed, in the appendix we show
that for a circular hole pocket and a single elliptical electron pocket
the error in the vector mean free path is of \emph{fourth} order in the eccentricity of the
electron pocket. If we henceforth neglect this shift, i.e., let
$\theta_\nu\approx\theta_{\nu-1}$ for all $\nu$, we incur an error that
is small for the moderate ellipticities of the electron Fermi pockets of
the pnictides. In the following section
we shall see that this convenient approximation generally compares
well with the full numerical solution of Eqs.\ (\ref{eq:03a}) and (\ref{eq:03b}).

Accordingly setting $\bar{\theta}_{s'}=\theta$ in Eq.\ (\ref{eq:05}), the
vector mean free paths for different $\theta$ decouple and we obtain
\begin{eqnarray}
\vl_{h\theta} &=& \frac{\vl_{h\theta}^{(0)}+\frac{1}{2}\, \big(a_{h\theta}^{e1}
  \vl_{e1\theta}^{(0)} + a_{h\theta}^{e2}\vl_{e2\theta}^{(0)}\big)}
  {1-\frac{1}{2}\,
  \big(a_{h\theta}^{e1}a_{e1\theta}^{h}+a_{h\theta}^{e2}a_{e2\theta}^{h}\big)},
\label{eq:07a} \\
 \vl_{e1\theta} &=& \vl_{e1\theta}^{(0)} + a_{e1\theta}^{h}\,\vl_{h\theta},
\label{eq:07b} \\
 \vl_{e2\theta} &=& \vl_{e2\theta}^{(0)} + a_{e2\theta}^{h}\,\vl_{h\theta}.
\label{eq:07c}
\end{eqnarray}
Results for the magnetic part $\vdl_{s\theta}$ can be found analogously by
replacing $\vl$ by $\vdl$ and $\vl^{(0)}$ by
\begin{equation}
\tau_{s\theta}\, \eta_s\,
  \frac{eB}{\pi\hbar^2}\, \frac{1}{N_{s\theta}}\,
  \frac{\partial\vl_{s\theta}}{\partial\theta} ,
\end{equation}
cf.\ Eq.\ (\ref{eq:03b}).
Since the anisotropy parameters $a_{s\theta}^{s'}$ are the only parameters in
the solution, apart from the RTA vector mean free paths, we will refer to these
expressions as the \emph{anisotropy approximation} (AA).
Clearly, for $a_{s\theta}^{s'}\ne0$ the vector mean free paths
involve the RTA solutions of all three Fermi pockets. This coupling
between the pockets becomes stronger
for larger anisotropy parameters. Additionally, the denominator in Eq.\
(\ref{eq:07a}), which appears in all results, provides a factor that is larger
than unity. In the anisotropic limit, $a_{s\theta}^{s'}\rightarrow1$, the vector
mean free paths $\vl_{s\theta}$ of all three pockets at a certain angle
$\theta$ become equal and diverge. Thus, for strong scattering anisotropy the vector
mean free path of the minority carriers must be \emph{inverted} relative to the
RTA result $\vl_{s\theta}^{(0)} \propto \mathbf{v}_{s\theta}$.

Semiclassically, we can interpret our results as follows. The solution to the
Boltzmann equation describes a non-equilibrium stationary state in which the
acceleration of the electrons due to external forces is balanced by scattering.
The vector mean free path of state $|s,\theta\rangle$ can be understood as
the displacement that an electron suffers until its velocity $\vv_{s\theta}$ is
randomized by scattering. The lifetime $\tau_{s\theta}$ is the mean time
between two scattering events. If the scattering is isotropic the velocity is
randomized after a single scattering event and the vector mean free path thus
reads $\tau_{s\theta}\vv_{s\theta}\equiv\vl^{(0)}_{s\theta}$. On the other hand,
anisotropic scattering only partially randomizes the velocity so
that the effective relaxation time exceeds the lifetime $\tau_{s\theta}$,
giving rise to multiple scattering during the relaxation, see Fig.\
\ref{fig:figure2}. The enhancement by denominator in Eq.\ (\ref{eq:07a})
accounts for this fact. In the extreme limit of $a_{s\theta}^{s'}\rightarrow1$,
the factor diverges, indicating that the velocities cannot relax at all and the
vector mean free paths become infinite.

This physical picture also applies to the case of two \emph{circular} Fermi
pockets considered in Refs.\ \onlinecite{Fanfarillo2012} and
\onlinecite{Breitkreiz2013}.
Because of rotational symmetry,
the vector mean free path is parallel to the velocity in that case,
and the AA becomes exact. This permits
a simple description in terms of transport times. However, we are here 
concerned with \emph{noncircular} Fermi pockets, which means that the
vector mean free 
path is generally not parallel to the velocity. The common feature is
that strong anisotropic scattering forces the vector mean free path of electron
and hole pockets at $\theta$ to point in the same direction, which is set by
the majority carriers. In the relevant parameter range for our model, we will
find that the direction is set by the electrons since there are two electron
pockets. A change of the dominant carrier type can only be achieved by
strong hole doping.

\section{Numerical results}
\label{NR}

To obtain quantitative results without further approximations beyond the
choice of the model and the semiclassical transport theory, we calculate the
scattering rate given in Eq.\ (\ref{eq:02}) by numerical integration.
Furthermore, we discretize the polar angle $\theta$, choosing $160$ sites on
each Fermi pocket. We have checked that taking
more points does not significantly change the results. The lifetimes, Eq.\
(\ref{eq:lt}), and the anisotropy parameters, Eq.\ (\ref{eq:06}), are obtained
by summation over the discrete sites. Finally, Eqs.\ (\ref{eq:03a}) and
(\ref{eq:03b}) are solved numerically by matrix inversion.
The numerical results will be compared to the AA, which is given by inserting
the lifetimes and the anisotropy parameters into Eqs.\
(\ref{eq:07a})--(\ref{eq:07c}).

\subsection{Scattering rate}

\begin{figure}[t]
\includegraphics[width=\columnwidth]{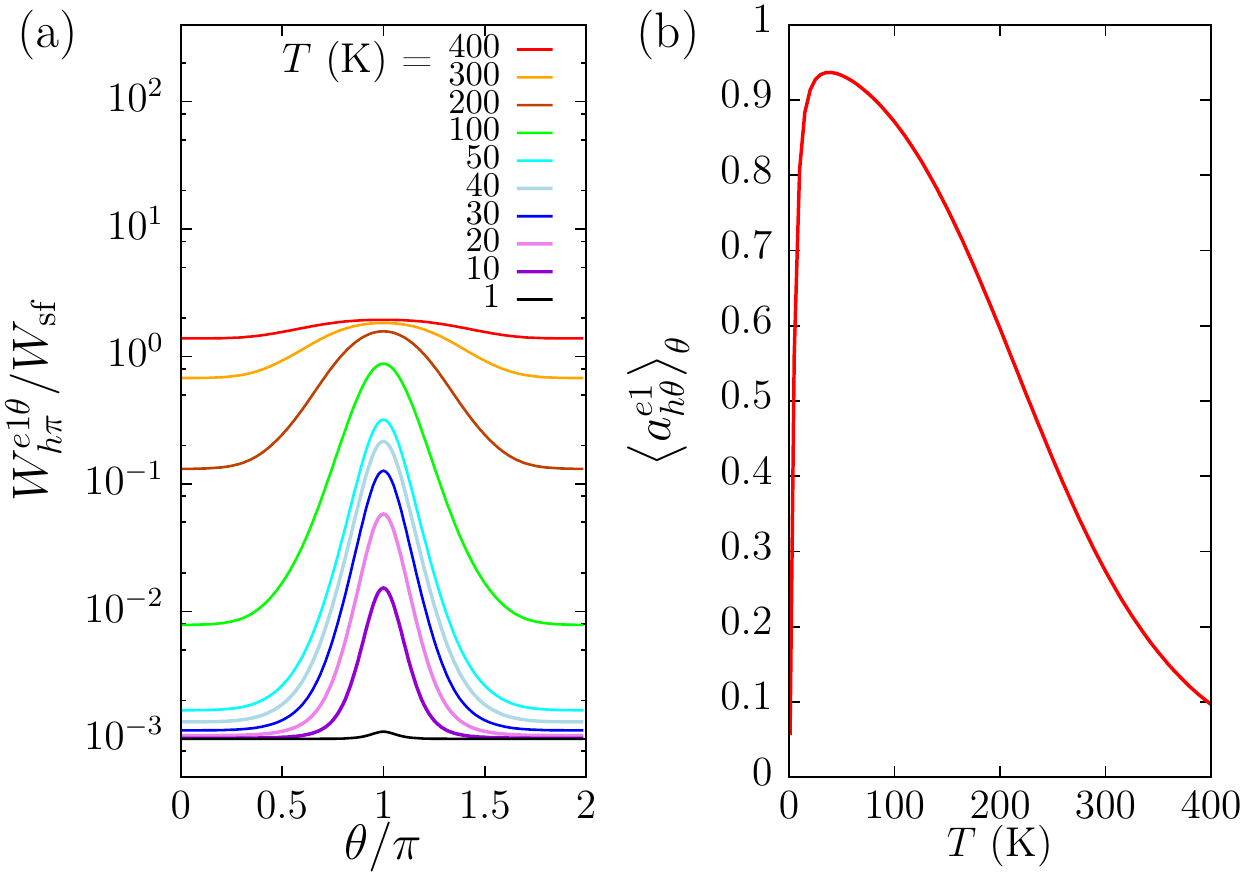}
\caption{(Color online) (a) Scattering
rate at different temperatures for an
electron in state $|h,\pi\rangle$ on the hole pocket to scatter to the state
$|e1,\theta\rangle$ on the electron pocket $e1$, as a function of
the final-state angle $\theta$. (b) Temperature dependence of the anisotropy
parameter averaged over all angles $\theta$ for the scattering shown in panel
(a). The parameters have been set to $\xi=1$, $n=2.08$,
and $W_{\text{imp}}/W_{\text{sf}}=10^{-3}$.}
\label{fig:figure7}
\end{figure}

Figure \ref{fig:figure7}(a) shows the temperature dependence of the scattering
rate for $\xi=1$ in Eq.\ (\ref{eek}) and $W_{\text{imp}}/W_{\text{sf}}=10^{-3}$.
While at high temperatures the scattering rate is isotropic, at
lower temperatures a peak due to spin fluctuations develops
corresponding to scattering vectors close to $\mathbf{Q}_{e1}$ or
$\mathbf{Q}_{e2}$. The peak becomes sharper as the temperature is lowered so
that the scattering anisotropy increases. At very low
temperatures spin fluctuations freeze 
out and only the isotropic impurity scattering remains so that the anisotropy
vanishes again. In Fig.\ \ref{fig:figure7}(b) we plot the anisotropy
parameter corresponding to the scattering rate shown in
Fig.\ \ref{fig:figure7}(a), averaged over the Fermi 
pocket. It clearly exhibits the increase for decreasing temperature and the
final sharp downturn at very low temperatures.
Note that in real pnictides, this low-temperature behavior will in
most cases be preempted by antiferromagnetic or superconducting order,
which are not described by our model spin susceptibility.

\subsection{ Hot-spot picture}
\label{CC}

\begin{figure*}[t]
\includegraphics[width=\textwidth]{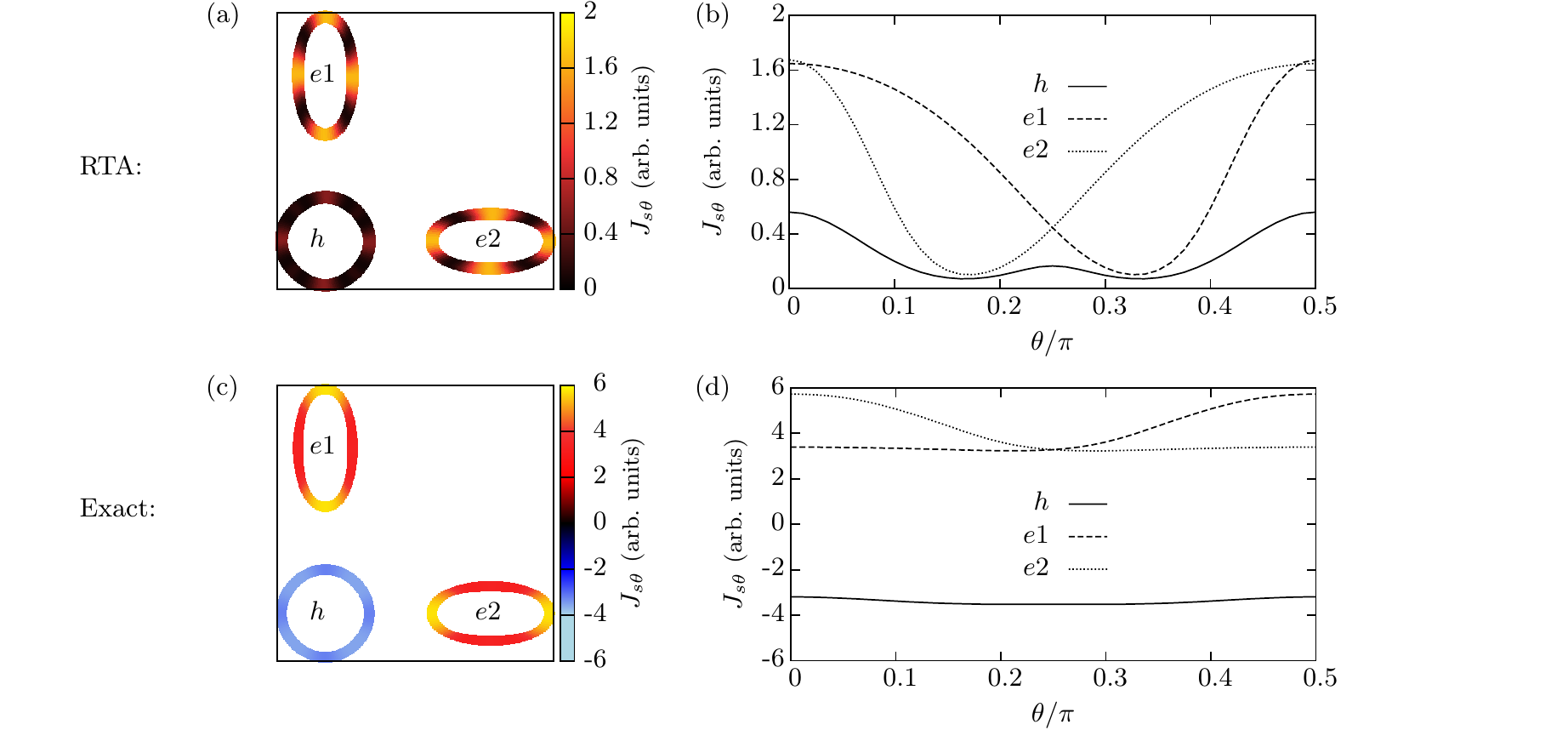}
\caption{(Color online) Contributions to the current of states on the Fermi
surface. Panels (a) and (b) show the RTA result as a color map and as a line
plot along a quarter of the Fermi pockets, respectively. Note the reduction of
the current contributions in the hot regions. Panels (c) and (d) show the same
for the full numerical results. The large anisotropy
leads to negative current contributions from the hole pocket. No signatures of
hot spots are apparent. The parameters are $\xi=2.5$, $n=2.08$,
$T=1\,\mathrm{K}$, and $W_{\text{imp}}/W_{\text{sf}}=0$.}
\label{fig:figure3}
\end{figure*}

In this subsection we explore how different parts of the Fermi pockets
contribute to the transport. In particular, we want to find out to what extent
the concept of hot and cold regions is applicable. Choosing $T=1\,\mathrm{K}$
and $W_{\text{imp}}/W_{\text{sf}}=0$, we focus on the regime of strong spin
fluctuations with strong scattering anisotropy, where the difference between the RTA and
the full result is the most striking.

The current parallel to the electric field is given by Eq.\
(\ref{eq:current12}).
The state-resolved current contributions $J_{s\theta}$ depend on the direction
of the electric field due to the noncircular Fermi pockets but we are
here not interested in this dependence and therefore average $J_{s\theta}$
over all directions of the electric
field in the \textit{xy}-plane. For $\mathbf{B}=0$ this gives
\begin{equation}
J_{s\theta} \equiv e^2N_{s\theta} \,
  \frac{v_{s\theta}^{x}\Lambda_{s\theta}^{x}+v_{s\theta}^{y}
  \Lambda_{s\theta}^{y}}{2}\, E.
\label{eq:current}
\end{equation}
Figure \ref{fig:figure3} shows the contributions $J_{s\theta}$ resulting from
the RTA as well as from the full numerical calculation. The two are completely
different.
Most prominently, the hot-spot
picture\cite{Fernandes2011,Eom2012,Arsenievic2013,Blomberg2013} is no longer
valid if forward-scattering corrections are taken into account. As discussed
above, the scattering off spin fluctuations is strongest in the hot regions
since the spin susceptibility is peaked at $\mathbf{Q}_{e1}$ and
$\mathbf{Q}_{e2}$, see Fig.\ \ref{fig:figure1}.
Thus the lifetimes are shorter and the RTA vector mean free paths given in Eqs.\
(\ref{eq:04a}) and (\ref{eq:04b}) are smaller. This is indeed reflected by the
suppressed current contributions in the hot regions shown in Figs.\
\ref{fig:figure3}(a) and \ref{fig:figure3}(b).
However, no signatures of hot regions are seen in the full results in Figs.\
\ref{fig:figure3}(c) and \ref{fig:figure3}(d).
This is due to the anisotropy of the scattering rate. In the hot regions, the
anisotropy $a_{s\theta}^{s'}$ is enhanced relative to the cold
  regions and, according to Eqs.\ 
(\ref{eq:07a})--(\ref{eq:07c}), this leads to an enhancement
of the vector mean free path, as was discussed in section
\ref{AA}. Thus the reduction of the lifetimes is
compensated by the enhanced scattering anisotropy and the
contribution of the hot regions to the current is comparable to that
of other parts of the Fermi pockets, i.e., the short-circuiting
of the hot spots does not occur. This insight is a central result
of our work.

Figure \ref{fig:figure3} also shows that
the holes contribute negatively to the total
current in the full calculation. In the semiclassical picture, this means that
the holes drift
in the same direction as the electrons. The insights gained in section
\ref{AA} illuminate this behavior: For the set of parameters chosen in Fig.\
\ref{fig:figure3}, the scattering anisotropy averaged over all Fermi states is
close to unity, $\left\langle a\right\rangle_\theta=0.96$. As discussed
in section
\ref{AA}, such a huge anisotropy leads to an effective relaxation time that is
much longer than the lifetime. In effect, during the relaxation,
an electron initially on the hole Fermi pocket scatters multiple times
between states on the hole pocket and states on the electron Fermi
pockets, which have nearly opposite velocity. Since there are more states on
the electron pockets than on the hole pocket, the
electron spends the larger part of the time on the electron pockets. The
electron thus on average drifts in the opposite direction to what one would get
if it stayed on the hole pocket. The RTA is not sensitive to the inversion
of the velocity upon interpocket scattering and thus cannot account for this
effect.

\subsection{Transport coefficients}

The transport coefficients can be obtained from the vector mean free
paths. The conductivity tensor is given in Eq.\ (\ref{eq:sigma}),
while the thermoelectric tensor reads\cite{Beenakker2011}
\begin{equation}
\alpha^{ij} = -\frac{\pi^2k_B^2T}{3e}\,
  \frac{\partial\sigma^{ij}}{\partial\mu} .
\label{alpha}
\end{equation}
We will focus on the resistivity
\begin{equation}
\rho=\frac{1}{\sigma^{xx}} ,
\label{rho}
\end{equation}
the Hall coefficient,
\begin{equation}
R_H=\frac{\sigma^{xy}}{(\sigma^{xx})^2B} ,
\label{RH}
\end{equation}
the Seebeck coefficient (thermopower),
\begin{equation}
S=-\frac{\alpha^{xx}}{\sigma^{xx}} ,
\label{Seebeck}
\end{equation}
and the Nernst coefficient,
\begin{equation}
{\cal N}=
  \frac{\sigma^{xy}\alpha^{xx}-\sigma^{xx}\alpha^{xy}}{(\sigma^{xx})^2B} .
\label{Nernst}
\end{equation}
We give the resistivity in units of
\begin{equation}
\rho_0 \equiv \frac{\hbar}{e^2}\, \frac{\hbar W_{\text{sf}}}{{\cal V}_0}\times
  10^{-2}\,(\mathrm{eV})^2 ,
\end{equation}
where ${\cal V}_0$ is the volume of the unit cell, and the Nernst coefficient in
units of
\begin{equation}
{\cal N}_0 \equiv \frac{{\cal V}_0}{e\rho_0}\times 10^{-5}\, \mathrm{V/K} .
\end{equation}
For the scattering strength ratio we choose in the following
$W_{\text{imp}}/W_{\text{sf}}=10^{-3}$.

\subsubsection{Comparison of approximations}

\begin{figure*}[t]
\includegraphics[width=\textwidth]{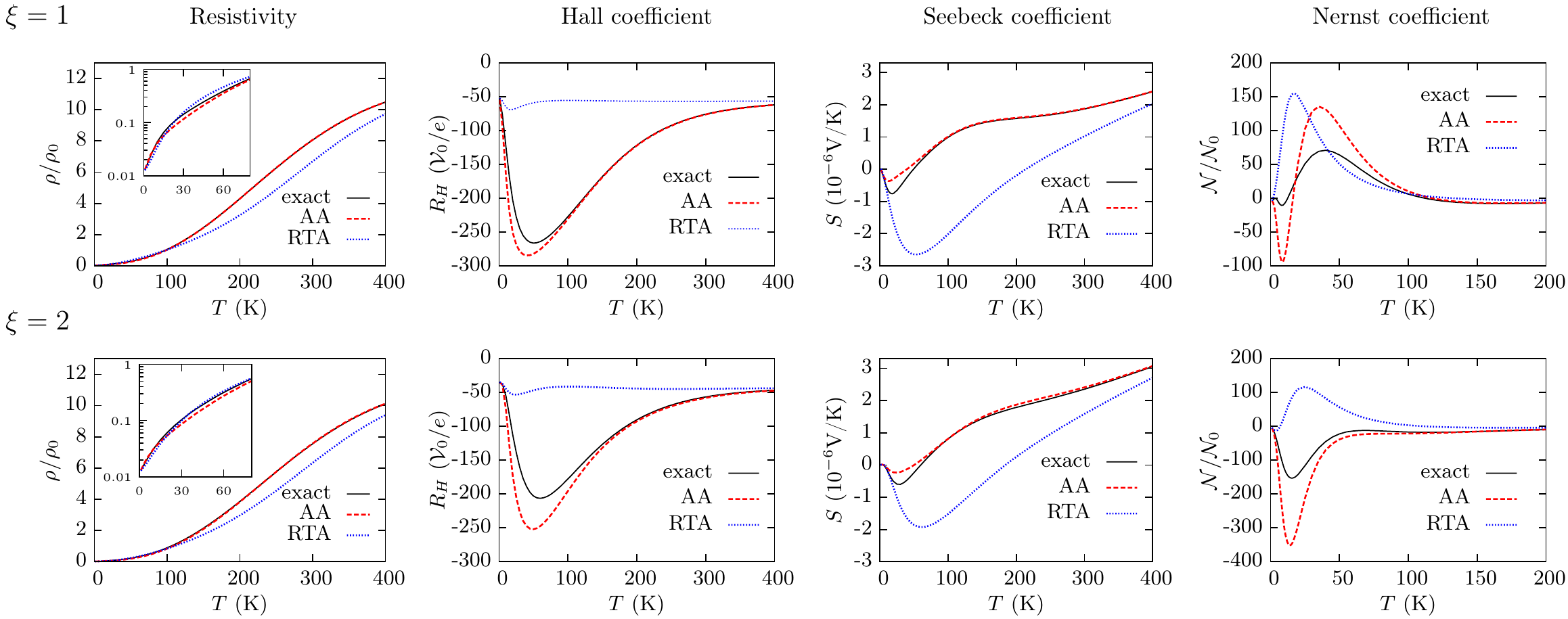}
\caption{(Color online) Temperature dependence of transport coefficients for
filling $n=2.05$ and ellipticity parameters $\xi=1$ and $\xi=2$, calculated with
three different methods: Numerically (``exact''), semianalytically within
the anisotropy approximation (``AA'') of Eqs.\ (\ref{eq:07a})--(\ref{eq:07c}),
and within the RTA, Eqs.\ (\ref{eq:04a}) and (\ref{eq:04b}).}
\label{fig:figure4}
\end{figure*}

Figure \ref{fig:figure4} shows the temperature dependence of the
transport coefficients, comparing the full numerical result with the RTA and the
AA. We see that the RTA results tend to coincide with the full calculation only
at very high and very low temperatures, where the scattering
is nearly isotropic, see Fig.\ \ref{fig:figure7}. In the temperature range with
strong anisotropy (20--150\,K) the deviations from the RTA are huge.
On the other hand, the AA shows qualitative agreement with the full results
over all temperatures and for both ellipticities. The agreement is
even quantitative for the resistivity. It is the worst for the Nernst
coefficient ${\cal N}$ but even here the positive and negative extrema in ${\cal
N}$ are predicted by the AA close to the correct temperatures. For
$\xi=1$ the AA is slightly better than for $\xi=2$ since the former value leads
to less eccentric electron pockets. The close agreement between
the AA and the full numerical results shows that the transport
behavior does not sensitively depend on the precise details of the
anisotropic scattering, and thus justifies our use of the approximate
susceptibility in Eq.~(\ref{eq:01}).

Both the RTA and the full results show strong temperature dependence.
For the RTA, this can be traced back to the nontrivial geometry of the Fermi
pockets leading to the hot-spot structure for high scattering anisotropies.
However, as discussed in subsection \ref{CC}, forward-scattering corrections
invalidate the hot-spot picture for strong anisotropies. The temperature
dependence of the RTA results thus stems from the wrong origin.
The true temperature dependence can be understood on the basis of the AA,
which gives qualitatively correct results. Here, it is due to the
strong temperature dependence of the anisotropy parameters $a_{s\theta}^{s'}$
shown in Fig.\ \ref{fig:figure7}(b), i.e., it relies on the \emph{corrections}
to the RTA in Eqs.\ (\ref{eq:03a}) and (\ref{eq:03b}) as well as
(\ref{eq:07a})--(\ref{eq:07c}).

The differences between the RTA and the full results for the resistivity and the
Hall coefficient are consistent with the predictions of Ref.\
\onlinecite{Breitkreiz2013} for two circular Fermi pockets.
In the resistivity, we note that the expected
enhancement and reduction for high and low scattering anisotropies,
respectively, lead to a more pronounced change of slope compared to the RTA.
Although the difference between the RTA and the full resistivity is
relatively small compared to the large corrections to the electron and hole
contributions shown in Fig.\ \ref{fig:figure3}, these corrections have
opposite signs and thus partially compensate each other, as already found for
circular Fermi pockets in Refs.\ \onlinecite{Fanfarillo2012,Breitkreiz2013}.
The predicted enhancement of the Hall coefficient is also
present.~\cite{Fanfarillo2012,Breitkreiz2013} However, the extremum of the Hall
coefficient in Fig.\
\ref{fig:figure4} is due to the maximum in the scattering anisotropy (cf.\ Fig.\
\ref{fig:figure7}) and is thus of different origin than in Ref.\
\onlinecite{Breitkreiz2013}, where  a maximum in the Hall coefficient was
predicted for the case that the anisotropy crosses a characteristic anisotropy
level at which the mobilities of holes and electrons are of equal magnitude but
opposite sign. We do not see any signatures of
such a crossing in the present results.
For the thermoelectric effects, Fig.\ \ref{fig:figure4} shows that the RTA
results are even qualitatively incorrect, with the Seebeck and Nernst
coefficients showing the wrong sign in the temperature range with strong
scattering anisotropy. According to Eqs.\ (\ref{alpha}) and (\ref{Seebeck}), the Seebeck
coefficient $S$ is proportional to $\partial\ln\sigma^{xx}/\partial\mu
= - \partial\ln\rho/\partial\mu$. In the
RTA, it stems from the shift of the hot spots with the chemical potential,
i.e., with doping. In the full results and the AA, it is instead due to the
change in the anisotropy parameters $a_{s\theta}^{s'}$ with the chemical
potential. Figure \ref{fig:figure4} shows that for the chosen parameters, the
two effects contribute to $S$ with opposite sign.
The full results for the Nernst coefficient ${\cal N}$ change sign between the
ellipticities $\xi=1$ and $\xi=2$. This effect is missed by the RTA. We return
to the Nernst coefficient below.

Qualitative differences between the RTA and the full solution of the
Boltzmann equation have also been reported for single-band
cuprate models with strongly anisotropic scattering.\cite{Beenakker2011, Kontani2008} 
The physics discussed here, including the inverted vector mean free path of
minority carriers, rely on the presence of multiple bands and Fermi pockets,
though.

\subsubsection{Doping dependence}

\begin{figure*}[t]
\includegraphics[width=\textwidth]{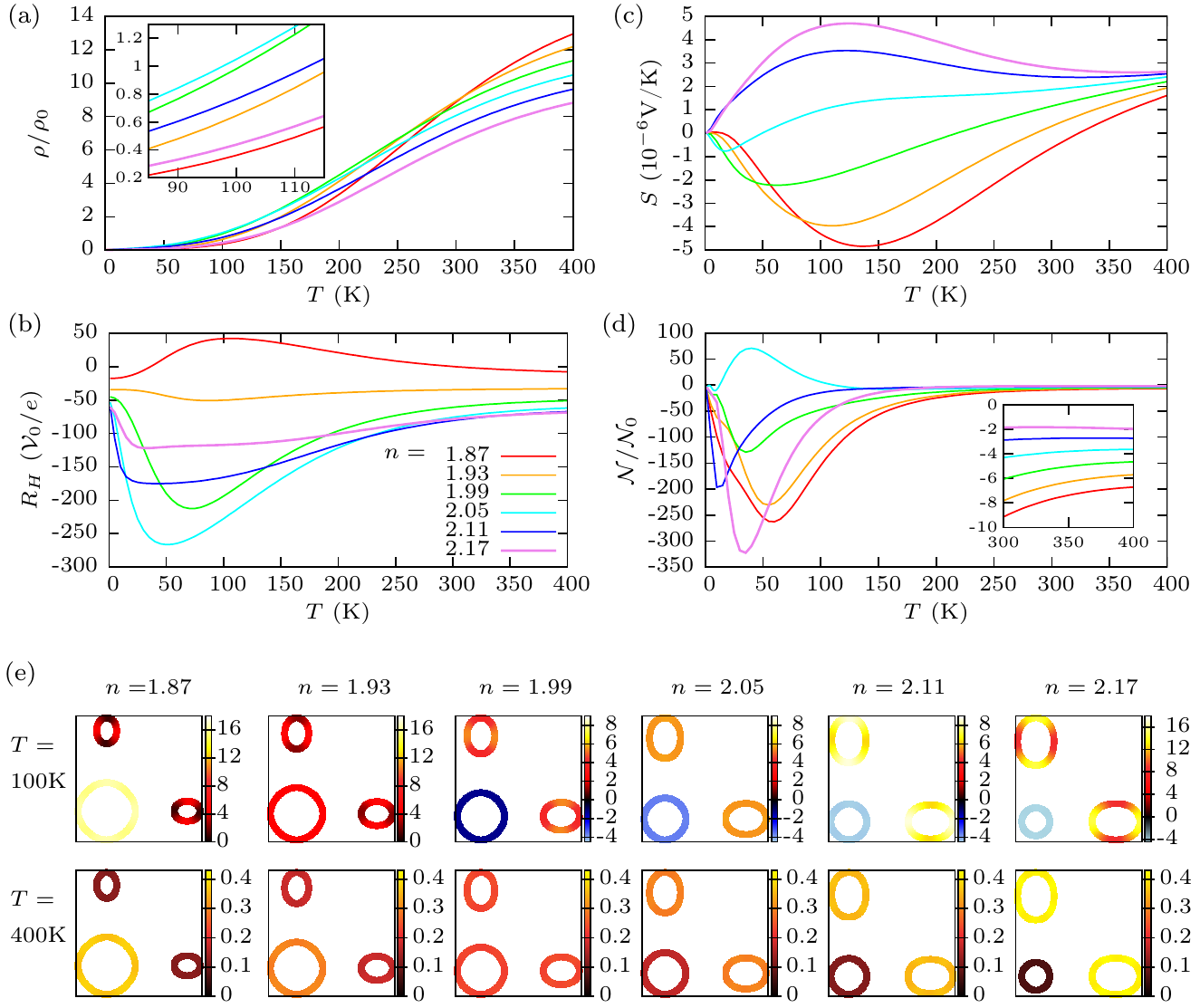}
\caption{(Color online) (a) Resistivity, (b) Hall
coefficient, (c) Seebeck coefficient, and (d) Nernst coefficient
as functions of temperature for different fillings $n$. (e)
State-resolved current contributions for $T=100\,\mathrm{K}$ and
$T=400\,\mathrm{K}$ for all fillings considered. Note the different color
scales.}
\label{fig:figure5}
\end{figure*}

We now turn to the doping dependence of the transport coefficients. Figures
\ref{fig:figure5}(a)--\ref{fig:figure5}(d) show the full solutions at different
fillings, while Fig.\ \ref{fig:figure5}(e) shows the current contributions of
states on
the Fermi surfaces at the two temperatures $T=100\,\mathrm{K}$ and
$T=400\,\mathrm{K}$ with strong and weak scattering anisotropy, respectively.
Note that the current contributions from the hole pocket are negative for
$T=100\,\mathrm{K}$ and $n\gtrsim 1.99$, i.e., towards the electron-doped side.
On the hole-doped side, the scattering is more isotropic due to the large
discrepancy in size between the electron and hole pockets.

At high temperatures, the transport coefficients all show a smooth doping
dependence resulting from the change in the Fermi surfaces and velocities in the
presence of mostly isotropic scattering. In
the intermediate temperature range, where anisotropic scattering is strong,
this is overlaid by nontrivial doping dependence due to the forward-scattering
corrections.

The resistivity around $T\approx 100\,\mathrm{K}$ is largest for
intermediate fillings, for which the Fermi pockets are well nested. This is
because the narrow peaks in the spin susceptibilities at $\mathbf{Q}_{e1}$ and
$\mathbf{Q}_{e2}$ lead to efficient scattering only for nested Fermi pockets.
The inefficiency of anisotropic scattering for small and large $n$ causes a
rapid decrease in the resistivity with doping, as shown in the inset in Fig.\
\ref{fig:figure5}(a). Note that the \emph{relative} change in $\rho$ with
doping is much larger here than at high temperatures. Since the Seebeck
coefficient $S$ is proportional to
$\partial\ln\sigma^{xx}/\partial\mu = -\partial\ln\rho/\partial\mu= -\rho^{-1}
\partial\rho/\partial\mu$, it is sensitive to this relative change in
$\rho$ with $\mu$ or $n$ and is, therefore, strongly enhanced in the
intermediate temperature range with strong scattering anisotropy, as Fig.\
\ref{fig:figure5}(c) clearly shows.

For the Hall coefficient $R_H$, Fig.\ \ref{fig:figure5}(b), one would naively
expect the largest and smallest values for the most strongly hole-doped and
electron-doped
cases, respectively, since electrons and holes contribute with opposite signs.
This is indeed the case at $T\approx 400\,\mathrm{K}$, where the scattering is
nearly isotropic and no negative current contributions occur.
At $T\approx 100\,\mathrm{K}$, however, Fig.\ \ref{fig:figure5}(b) shows a
strong negative enhancement of $R_H$ for intermediate filling.
According to Fig.\ \ref{fig:figure5}(e), the contribution of the holes to the
total current is negative in this range. In the semiclassical picture this means
that the holes drift in the same direction as the electrons, reducing the
charge current. Irrespective of that, the magnetic field deflects the holes and
the electrons in the same direction. Hence, the inverted sign of the hole
contribution reduces the charge current without changing the Hall voltage.
This gives rise to an enhancement of the Hall coefficient defined as the Hall
voltage relative to the charge current.

The Nernst coefficient ${\cal N}$ plotted in Fig.\ \ref{fig:figure5}(d) is
highly sensitive to small doping changes and also, as is evident from Fig.\
\ref{fig:figure4}, to changes in the band parameters. Equations
(\ref{alpha})--(\ref{RH}) and (\ref{Nernst}) show that
\begin{equation}
{\cal N} = \frac{3e}{\pi^2k_B^2T}\, \frac{\partial}{\partial\mu}\,
  \frac{R_H}{\rho}
  = \frac{3e}{\pi^2k_B^2T}\, \frac{\partial n}{\partial\mu}\,
  \frac{\partial}{\partial n}\, \frac{R_H}{\rho} .
\end{equation}
The Nernst coefficient is thus sensitive to the nonmonotonic doping dependence
of both $\rho$ and $R_H$. For the cases we have considered, the contributions
from $\rho$ and $R_H$ usually counteract each other. The complicated behavior of
${\cal N}$, for example the different sign of ${\cal N}$ for $n=2.05$ compared
to the other fillings, is thus due to the quantitative competition of the doping
dependences of $\rho$ and $R_H$ and not to any clear qualitative features in the
Fermi surfaces or the scattering. This suggests that the other coefficients might
be more advantageous as probes of the electronic system. However, the detailed
comparison of experimental transport coefficients and calculations for
realistic models remains work for the future.

\section{Conclusions}
\label{Con}

We have studied transport in a two-band model relevant for the iron pnictides,
using the semiclassical Boltzmann equation. Forward-scattering corrections
due to anisotropic interband scattering off spin fluctuations have been
included. Spin fluctuations have been described by a phenomenological
Millis-Monien-Pines susceptibility,\cite{Millis1990} with temperature-dependent
parameters chosen based on neutron-scattering results for the
pnictides.\cite{Inosov2010} Our analytical and numerical investigations show
that the anisotropic scattering gives rise to unusual transport behavior.
Most surprisingly, the hot spots are \emph{not} short-circuited
by the cold regions of the Fermi 
pockets even for very strong scattering. The enhanced scattering rate in the hot
regions indeed leads to a short lifetime there, but this effect is
balanced by the enhanced vector mean free path due to the anisotropic
scattering. This breakdown of the concept of hot and cold regions is not found
in a simple RTA neglecting forward-scattering corrections.

The nearly isotropic contribution of states around the Fermi pocket to
the transport, even for strongly elliptical electron pockets, justifies
the discussion of transport in terms of isotropic mobilities for each
pocket. However, as discussed for the case of circular
pockets,\cite{Breitkreiz2013,Fanfarillo2012} the mobility of the minority
carriers can turn negative in the regime of highly anisotropic scattering. In
the present work, negative mobility corresponds to inverted vector mean
free paths and the resulting negative current contributions.

The contribution of hot spots to the transport and the occurrence of
negative currents  are the main features that distinguish the transport
properties of pnictides from previously considered one-band systems
with similarly anisotropic scattering.
In this work, we have presented unusual temperature and doping 
dependences of various transport coefficients.
Beyond this, negative current contributions can also lead to a negative
magnetoresistance.\cite{Breitkreiz2013} However, the present model with two
electron pockets and one hole pocket does not show negative
magnetoresistance in the considered parameter range. Calculations of transport
coefficients for more realistic pnictide models are desirable to allow
quantitative predictions.

\acknowledgments

Financial support by the Deutsche Forschungsgemeinschaft through Research
Training Group GRK 1621 is gratefully acknowledged. The authors thank
J. Schmiedt, H. Kontani, and O. Kashuba for useful discussions.

\appendix

\section*{Appendix: Discussion of the angular shift}

\begin{figure}[t]
\includegraphics[width=\columnwidth]{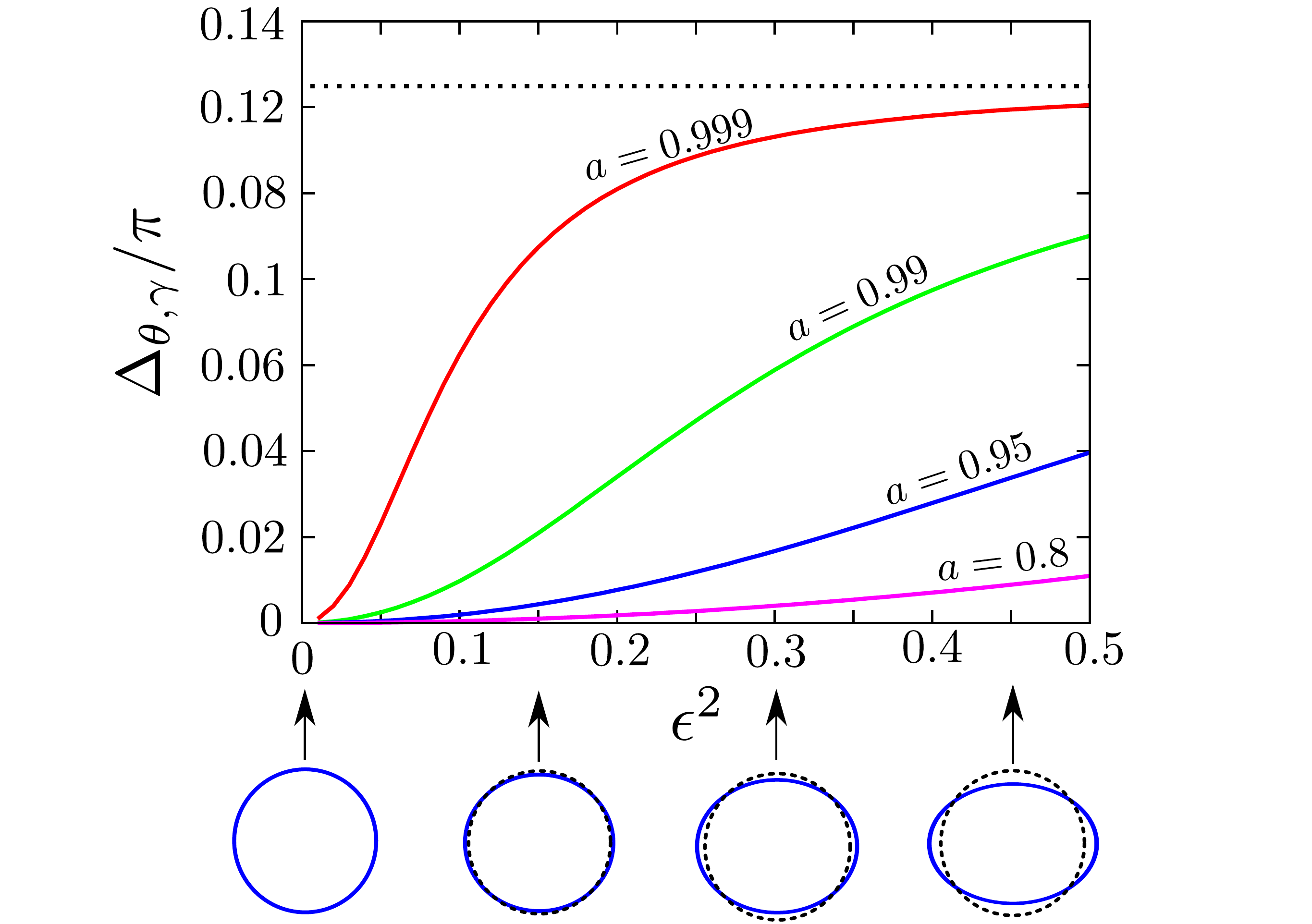}
\caption{(Color online) Effective angular shift $\Delta_{\theta,\gamma}$ for
the state $\theta=\pi/8$ as a function of $\epsilon^2$ for different uniform
anisotropies $a$. $\epsilon$ represents the eccentricity of the electron
pocket. The shift is directed towards the hot spot at $\theta=\pi/4$,
hence the maximal value of $\Delta_{\theta,\gamma}/\pi$ is $1/8$. The
insets indicate the shape of the electron pocket (solid line) corresponding to
various values of $\epsilon^2$. }
\label{fig:shift}
\end{figure}

As discussed in the main text, the vector mean free path of a state
$|s,\theta\rangle$ can be written as a power series in the anisotropy
parameter, where the term of order $n$ contains the RTA vector mean free path
of a state reached by $n$ hopping events towards the closest hot spot. We have
argued that the angular shift towards the hot spot is a small effect for
the vector mean free path for
realistic ellipticities of the electron pockets and have therefore ignored it
above. We here explore this effect analytically within a simple model. To get an
estimate for the upper limit of the correction to the vector mean free path, it
is sufficient to consider only a \emph{single} electron pocket. Our simple model
consists of a circular hole Fermi pocket with the Fermi wave number $k$ and
an elliptical electron Fermi pocket described by the semi-major and
semi-minor axis $k_a=k(1-\epsilon^2)^{-1/4}$ and $k_b=k(1-\epsilon^2)^{1/4}$,
respectively, where $\epsilon$ is the eccentricity of the ellipse. To focus on
the shift effect we assume constant anisotropy, $a_{s\theta}^{s'}=a$. For
two Fermi pockets and constant anisotropy, Eq.\ (\ref{eq:05}) takes the
form
\begin{equation}
\vl_{s\theta}=\vl_{s\theta}^{(0)}+a\, \vl_{\bar{s}\bar{\theta}},
\label{eq:A01}
\end{equation}
where $\bar{h}=e$, $\bar{e}=h$, and the RTA solution $\vl^{(0)}$ is given by
Eq.\ (\ref{eq:04a}). Using simple trigonometry, we find that for the given geometry,
the difference between
$\bar{\theta}$ and $\theta$ to leading order in the eccentricity $\epsilon$
reads $\frac{\epsilon^4}{16}\sin 4\theta$. Iterating
Eq.\ (\ref{eq:A01}), we obtain the solution for the electron pocket as
\begin{equation}
\vl_{e\theta} = \sum_{n=0}^{\infty} a^{2n}\,
  \big(\vl^{(0)}_{e\theta_n}+a \vl^{(0)}_{h\theta_n}\big),
\label{eq:A02}
\end{equation}
with
\begin{equation}
\theta_n=\theta_{n-1} + \frac{\epsilon^4}{16}\, \sin 4\theta_{n-1} \quad
  \text{and}\quad \theta_0=\theta.
\label{eq:A03}
\end{equation}
The solution for the hole pocket follows immediately from Eqs.\ (\ref{eq:A01})
and (\ref{eq:A02}).

Replacing the discrete index $n$ by a continuous variable, we obtain
\begin{equation}
\vl_{e\theta} = -\frac{2\ln a}{1-a^{2}}
  \int_{0}^{\infty} dn\, a^{2n} \,
  \big(\vl_{e\theta_{n}}^{(0)}+a\vl_{h\theta_{n}}^{(0)}\big)+\mathbf{R},
\end{equation}
with a correction $\mathbf{R}$. By splitting the integration range into
intervals $[m,m+1]$ with integer $m$, one can easily show that
\begin{equation}
\left|\mathbf{R}\right| \le
  \sum_n a^{2n}\left| \big(\vl^{(0)}_{e\theta_n}+a\vl^{(0)}_{h\theta_n}\big)
  -\big(\vl^{(0)}_{e\theta_{n+1}}+a\vl^{(0)}_{h\theta_{n+1}}\big) \right|,
\end{equation}
which is obviously of higher order in $\epsilon^2$ because of Eq.\
(\ref{eq:A03}).
Substituting $n=4\ln(1+z)/\epsilon^4$ we obtain
\begin{equation}
\vl_{e\theta} = \frac{1}{1-a^{2}}
  \int_{0}^{\infty} dz\, \gamma\, \bigg(\frac{1}{1+z}\bigg)^{\!\gamma+1}
  \big(\vl_{e\theta(z)}^{(0)}+a\vl_{h\theta(z)}^{(0)}\big) ,
\label{eq:A04}
\end{equation}
with
\begin{equation}
\gamma \equiv 8\,\frac{\ln(1/a)}{\epsilon^4}
\label{eq:A07}
\end{equation}
and
\begin{equation}
\theta(z) \equiv \frac{1}{2}\arctan\left[(z+1)\tan2\theta\right].
\end{equation}
In the integral in Eq.\ (\ref{eq:A04}), the factor
$\gamma\,[1/(1+z)]^{\gamma+1}$ acts as a distribution function which
is normalized to unity and becomes a $\delta$-function in the limit of zero
ellipticity, i.e., for $\gamma\rightarrow\infty$. Hence, the largest shifts are
achieved for small values of $\gamma$, which, according to Eq.\
(\ref{eq:A07}), correspond to large scattering anisotropy and large ellipticity.

The shift also depends on the position on the Fermi pocket. There is
no shift at
the hot spots, $\theta=(2n-1)\,\pi/4$, and at the cold spots, $\theta=n\,\pi/2$.
The largest shift can be expected to occur between the hot and cold spots, in
the vicinity of $(2n-1)\,\pi/8$.

We can make further analytical progress by expanding the vector
$\big(\vl_{e\theta(z)}^{(0)}+a\vl_{h\theta(z)}^{(0)}\big)$ to linear order in
$\theta(z)$. This is best justified if the total angular shift is small, i.e.,
if we start with $\theta$ close to a hot spot. However, the total shift can
never be larger than $\pi/4$ so that the approximation always gives at least
qualitatively correct results for not excessive eccentricities.
Equation (\ref{eq:A04}) can then be written as
\begin{equation}
\vl_{e\theta} = \frac{1}{1-a^{2}}\,
  \big(\vl_{e,\theta+\Delta_{\theta,\gamma}}^{(0)}
  +a\vl_{h,\theta+\Delta_{\theta,\gamma}}^{(0)}\big),
\label{eq:A05}
\end{equation}
with the effective angular shift
\begin{eqnarray}
\Delta_{\theta,\gamma} &=& \int_{0}^{\infty} dz\,
   \gamma\, \Big(\frac{1}{1+z}\Big)^{\!\gamma+1}\theta(z)-\theta \nonumber \\
&\cong& \frac{\sin4\theta}{32}\, \frac{\epsilon^4}{\ln(1/a)}
  + \frac{\sin8\theta}{512}\,
  \bigg[ \frac{\epsilon^4}{\ln(1/a)} \bigg]^2
\nonumber \\
&& {} + \mathcal{O}\bigg(\bigg[
  \frac{\epsilon^4}{\ln(1/a)}\bigg]^3\bigg) .
\label{eq:A06}
\end{eqnarray}
By neglecting the shift, $\Delta_{\theta,\gamma}=0$, we would obtain the
analogue of Eqs.\ (\ref{eq:07a})--(\ref{eq:07c}) for the case of constant
anisotropy and a single electron pocket.

In Fig.\ \ref{fig:shift} we plot the angular shift at $\theta=\pi/8$ for
different anisotropies as a function of the eccentricity squared, $\epsilon^2$.
Realistic scattering anisotropies hardly exceed the value $a=0.95$, for which the shift is
small up to $\epsilon^2\approx 0.5$. Stronger ellipticities might, however,
lead to significant corrections.

\end{document}